\documentclass{article}
\usepackage[T1]{fontenc} 
\usepackage[utf8]{inputenc} 
\usepackage{amsmath,cite,url}
\usepackage{amssymb}
\usepackage{graphicx}
\usepackage[lbd]{ismir}
\usepackage{booktabs}
\usepackage{xcolor}
\usepackage{acronym}

\usepackage[firstpage]{draftwatermark}
\definecolor{lightgray}{rgb}{0.9,0.9,0.9}
\definecolor{darkgray}{rgb}{0.4,0.4,0.4}
\SetWatermarkFontSize{12pt}
\SetWatermarkScale{1.1}
\SetWatermarkAngle{90}
\SetWatermarkHorCenter{202mm}
\SetWatermarkVerCenter{170mm}
\SetWatermarkColor{darkgray}
\SetWatermarkText{Late-Breaking / Demo Session Extended Abstract, ISMIR 2025 Conference}

\acrodef{OT}{Optimal Transport}
\acrodef{CQT}{Constant-$Q$ Transform}
\acrodef{VQT}{Variable-$Q$ Transform}
\acrodef{DDSP}{Differentiable Digital Signal Processing}
\acrodef{STFT}{Short-Time Fourier Transform}
\acrodef{TF}{Time-Frequency}
\acrodef{PSD}{Power Spectral Densitie}

\newcommand{\xt}{\bold{\tilde{x}}}
\newcommand{\xtk}{\bold{\tilde{x}}^{(k)}}

\newcommand{\yt}{\tilde{y}}
\newcommand{\ytk}{\tilde{y}^{(k)}}

\newcommand{\WW}{\mathcal{W}}
\newcommand{\LL}[1]{\mathcal{L}_{\text{#1}}}

\newcommand{\alain}[1]{#1}
\newcommand{\be}[1]{#1}

\title{Translation-Equivariant Self-Supervised Learning\\for Pitch Estimation with Optimal Transport}

\multauthor
{Bernardo Torres*
\hspace{1cm} Alain Riou*
\hspace{1cm} Gaël Richard
\hspace{1cm}  Geoffroy Peeters}
{LTCI, Télécom-Paris, Institut Polytechnique de Paris, France\\
{\small *Equal contribution, correspondence to bernardo.torres@telecom-paris.fr}
}

\def\authorname{B. Torres, A. Riou, G. Richard, and G. Peeters.}

\begin{document}

\maketitle
\begin{abstract}

In this paper, we propose an \acl{OT} objective for learning one-dimensional translation-equivariant systems
\alain{and demonstrate its applicability to single pitch estimation.}
Our method provides a theoretically grounded, more numerically stable, and simpler alternative for training state-of-the-art self-supervised \alain{pitch estimators}. 
\end{abstract}
\section{Introduction}\label{sec:introduction}


Pitch estimation is a core task in audio analysis, long studied in the speech and Music Information Retrieval (MIR) communities~\cite{Noll1967}. It involves estimating the fundamental frequency of harmonic or quasi-harmonic signals, with traditional methods relying on signal processing techniques to extract harmonicity cues~\cite{YIN,pYIN,Lee2012}, or by matching the input spectrum to that of a synthetic waveform~\cite{SWIPE}.


\be{Recently, supervised deep learning approaches leveraging large annotated datasets (such as CREPE~\cite{CREPE}) have achieved impressive accuracy, but come with notable challenges.
In particular, labeling audio with the temporal precision needed for training (typically within a few milliseconds) is labor-intensive and prone to errors. While synthetic data can help alleviate this~\cite{MDBstemsynth}, supervised models remain sensitive to out-of-distribution scenarios—e.g., when trained on music and evaluated on speech~\cite{PENN}.}

To address these limitations, self-supervised learning has emerged as an efficient and scalable alternative~\cite{SPICE,PESTO,Cwitkowitz2024}. To create pitch information without any labels, a pair of audio frames are pitch-shifted artificially by a \emph{known} shift $k$ and a Siamese network~\cite{Siamese} is trained to be equivariant to such pitch-shift by optimizing a criterion parametrized by $k$.
In practice, a \ac{CQT}~\cite{CQT} is used as a frontend to the network since it maps frequencies to a log scale, where the frequency difference between harmonic components is independent of the fundamental frequency. As a result, pitch shifting operation for harmonic sounds is roughly equivalent to a simple translation.

This technique was originally proposed in SPICE~\cite{SPICE}, in which the pitch estimator is trained by minimizing an equivariance objective between scalar pitch predictions. Later, PESTO \cite{PESTO, riouPESTORealtimePitch2025} proposed an alternative equivariance criterion that operates on estimated pitch distributions, while also taking advantage of a lightweight transposition-equivariant architecture.

\alain{Moreover, equivariance to translations is not restricted to single pitch estimation, and further works leveraged similar principles for tasks such as}
multi-pitch \cite{Cwitkowitz2024}, tempo~\cite{Quinton2022,Gagnere2024}, tonality\cite{kongSTONESelfsupervisedTonality2024} and key\cite{kongSKEYSelfsupervisedLearning2025} estimation.\\

In this paper, we simplify the training of PESTO by
\alain{introducing}
an \ac{OT}~\cite{villaniOptimalTransportOld2009} inspired objective between pitch distributions. \ac{OT} provides a framework to compare distributions taking into account their horizontal displacement, and it has been used audio for frequency-domain comparison of audio signals \cite{cazellesWassersteinFourierDistanceStationary2020,flamaryOptimalSpectralTransportation2016, torresUnsupervisedHarmonicParameter2024}.  We evaluate our proposed objective in the same cross-evaluation framework as \cite{riouPESTORealtimePitch2025} and show it has encouraging results.



\section{Optimal Transport}

 Let $\mu = \sum^n_{i=1} \mathbf{a}_i \delta_{p_i}$ and $\nu = \sum^m_{j=1} \mathbf{b}_j \delta_{q_j}$ be discrete distributions with weights $\mathbf{a}_i$ at positions $p_i$ and weights $\mathbf{b}_j$  at positions $q_j$, where $\delta_{p}$ is the Dirac measure at location $p$. 
Let ($\mathbf{a}$, $\mathbf{b}$)  belong to the space of probability vectors, $i.e.$ $\mathbf{a} \in \Sigma_n$ and $\mathbf{b} \in \Sigma_m$, for $\Sigma_n = \left\{\mathbf{a} \in \mathbb{R}^n_+ ; \sum_{i=1}^n \mathbf{a}_i = 1 \right\}$. 

Given a cost function $c(p_i, q_j)$ representing the cost of transporting a unit of mass from locations $p_i$ to $q_j$, the \ac{OT} problem seeks to find the optimal path of moving $\mu$  to $\nu$ such that the total transportation cost  $\mathcal{L}_c(\mu, \nu)$, for ground cost function $c$, is minimal. When $c(p_i, q_j) = \left| p_i - q_j \right|^p$, $\mathcal{L}_c(\mu, \nu)^{\frac{1}{p}}$ is called the $p$-Wasserstein distance ($\mathcal{W}_p$).




\medskip

\noindent \textbf{Closed form for 1D Wasserstein distances}: Let $F_\mu $ : $\mathbb{R}$ $ \to $ $[0, 1]$ denote the cumulative distribution function (CDF)  of $\mu$: $ F_{\mu}(t) = \sum_{i=1}^n u(t - p_i) \mathbf{a}_i $, where $u(t)$ is the step function. Let additionally $F^{-1}_\mu$ : $[0, 1]$ $\to$ $\mathbb{R}$ be its pseudoinverse (the quantile function). For one dimensional distributions, $\mathcal{W}_p$ can be conveniently expressed in closed form by integrating $ \left| F^{-1}_{\mu}(r) - F^{-1}_{\nu}(r) \right|^p$ on the real line \cite{ peyreComputationalOptimalTransport2019}, and can be approximated by a discrete sum \cite{flamaryPOTPythonOptimal2021short}:

\vspace{-2em}
\begin{equation}
    \WW_p(\mu, \nu) =  \sum_{i=1}^n \left| F^{-1}_{\mu}(r_i) - F^{-1}_{\nu}(r_i) \right|^p (r_i - r_{i-1}).
    \label{eq:wasserstein_discrete}
\end{equation}


\pagebreak

\noindent \textbf{Wasserstein distance under translations}: A key property of the $2$-Wasserstein distance is its simple behavior under translations. For a distribution $\mu$ and its version $\mu_k$ translated by $k$, it follows that $\mathcal{W}_2(\mu, \mu_k) = |k|$ \cite{peyreComputationalOptimalTransport2019}.


This property is highly relevant to audio signal processing. As shown in \cite{cazellesWassersteinFourierDistanceStationary2020}, if we consider the Power Spectral Densities ($S_1, S_2$) of two signals where one is a frequency-modulated version of the other ($s_2(t) = s_1(t)e^{2i\pi\xi_0 t}$), $\mathcal{W}_2(S_1, S_2) = |\xi_0|$ (it measures precisely the frequency shift $\xi_0$). This property inspired  Spectral Optimal Transport (SOT) \cite{torresUnsupervisedHarmonicParameter2024} to employ its discrete sum approximation (Eq. \ref{eq:wasserstein_discrete}) as a loss function on magnitude spectrograms.

However, this does not directly apply to musical pitch shifting (i.e., transposition) when using a standard Fourier linear frequency representation. By working in the \ac{CQT} domain instead, the frequency coordinates are now logarithmically spaced, and $\mathcal{W}_2$ becomes a convex measure of the pitch shift (in semitones).


\section{Self-supervised equivariance loss based on optimal transport}\label{sec:ot-ssl}

We base this work on the improved version of PESTO \cite{riouPESTORealtimePitch2025}, in which the \ac{VQT} is used as the frontend for improved time localization. The encoder $\mathcal{F}$ is trained by optimizing a combination of three losses that favor \emph{equivariance} to pitch-shifting and \emph{invariance} to pitch-preserving transforms (noise, gain...), respectively~\cite{PESTO}:
\begin{equation}
    \label{eq:pesto}
    \LL{PESTO} = \underbrace{\lambda_{\text{equiv}} \LL{equiv} + \lambda_{\text{SCE}} \LL{SCE}}_{\text{equivariance}} + \underbrace{\lambda_{\text{inv}} \LL{inv}}_{\text{invariance}},
\end{equation}
where the $\lambda_*$ are dynamically updated based on losses' respective gradients, as detailed in \cite{riouPESTORealtimePitch2025}.

From pitch-shifted \ac{VQT} frames $\xt$ and $\xtk$, the equivariance loss terms are computed between  $\yt = \mathcal{F}(\xt)$ and $\ytk = \mathcal{F}(\xtk)$, $k$ being the translation/pitch-shift between $\xt$ and $\xtk$. $\LL{equiv}$ projects both distributions to scalar values using a vector of power series $(\alpha, \alpha^2, \dots, \alpha^n)^\top$ and compare the ratio of these scalars with $\alpha^k$. In addition, $\LL{SCE}$ forces $\yt$ and $\ytk$ to match up to a translation/pitch-shift of $k$ bins by minimizing their cross-entropy.

Here, we propose replacing the equivariance terms of Eq. \ref{eq:pesto} by the 1D Wasserstein distance (Eq. \ref{eq:wasserstein_discrete}, as implemented by \cite{torresUnsupervisedHarmonicParameter2024}) between $\yt$ and $\ytk$ (corrected by $k$). Let $\tau_k : \Sigma^n \to \Sigma^n$ denote the translation operator which shifts distribution $y$ by $k$ bins to the right\footnote{In practice, $\yt$ and $\ytk$ are padded with $k_{\max}$ zeros on both sides before applying the loss, then $\tau_k$ is implemented as a circular shift.}.
We train the model to minimize:
\begin{equation}
    \LL{OT}(\yt, \ytk, k) = \mathcal{W}_2 \left(\yt, \tau_{-k}(\ytk) \right).
\end{equation}
Contrary to $\LL{equiv}$ and $\LL{SCE}$,  this loss consists of a single term and has several desirable theoretical properties:

\begin{itemize}
    \item \textbf{Symmetry:} $\LL{OT}(y_1, y_2, k) = \LL{OT}(y_2, y_1, -k)$.
    \item \textbf{Invariance:} $\WW_p(y_1, y_2) = \WW_p(\tau_k(y_1), \tau_k(y_2))$.
    \item \alain{\textbf{Linear scaling under $\tau_k$:}} $\WW_2(y, \tau_k(y)) \propto |k|$.
    \item \textbf{Stability:} $\LL{OT}$ avoids the large floating-point powers $\alpha^i$ that can cause numerical instability in $\LL{equiv}$.
\end{itemize}

\vspace{-0.5em}

These properties are moderated in practice by the fidelity of the pitch-shifting approximation (i.e., \ac{VQT} cropping) and any boundary artifacts introduced by the shifting operation. The full objective becomes:
\begin{equation}
    \LL{PESTO-OT} = \lambda_{\text{OT}} \LL{OT} + \lambda_{\text{inv}} \LL{inv},
\end{equation}
with $\lambda_{\text{OT}}$ and $\lambda_{\text{inv}}$ updated as in \cite{riouPESTORealtimePitch2025}, based on gradients.

\section{Experiments}

Our evaluation protocol strictly follows the one from \cite{riouPESTORealtimePitch2025}. We use the same architecture, hyperparameters and training schedule as the configuration with \ac{VQT} parameter $\gamma=7$.
We train and evaluate our model on three datasets, covering singing voice (MIR-1K~\cite{MIR-1K}), music (MDB-stem-synth~\cite{MDBstemsynth}) and speech (PTDB \cite{pirkerPitchTrackingCorpus2011}). We refer to \cite{riouPESTORealtimePitch2025} for further details about the protocol and datasets.

We report in Table \ref{tab:results} cross evaluation Raw Pitch Accuracy (RPA) and Raw Chroma Accuracy (RCA) values for the considered datasets.
Despite no hyperparameter tuning, our method achieves competitive performances with the state-of-the-art PESTO baseline, especially when training on MIR-1K.
The biggest gap is observed when evaluating on MDB, which spans a wider pitch range.

These encouraging results, combined with the properties outlined in Section~\ref{sec:ot-ssl}, suggest the relevance of OT-based loss functions for equivariant SSL.

\begin{table}[]
    \centering
    \resizebox{\columnwidth}{!}{
    \begin{tabular}{lcccccccc}
        \toprule
         & \multicolumn{2}{c}{MIR-1K} &  & \multicolumn{2}{c}{MDB} &  & \multicolumn{2}{c}{PTDB} \\
        \cmidrule{2-3} \cmidrule{5-6} \cmidrule{8-9}
         & RPA & RCA &  & RPA & RCA &  & RPA & RCA \\
        \midrule
        PESTO~\cite{riouPESTORealtimePitch2025} & \\
        \; MIR-1K &
        97.7 & 98.0 &  & 94.8 & 95.9 &  & 87.7 & 90.3 \\
        \; MDB &
        94.6 & 96.1 &  & 97.0 & 97.1 &  & 88.3 & 89.9 \\
        \; PTDB &
        95.6 & 96.9 &  & 96.3 & 96.6 &  & 89.7 & 91.2 \\
        \midrule
        \multicolumn{5}{l}{PESTO-OT (Ours)} & \\
        \; MIR-1K &
        97.8 & 98.1 &  & 86.6 & 95.1 &  & 88.0 & 90.1 \\
        \; MDB &
        91.6 & 94.0 &  & 95.3 & 95.6 &  & 88.3 & 89.8 \\
        \; PTDB &
        86.4 & 92.9 &  & 93.7 & 94.5 &  & 89.0 & 90.8 \\
        
        \bottomrule
    \end{tabular}}
    \caption{Comparison of our model with the PESTO model\cite{riouPESTORealtimePitch2025} on different datasets. Rows and columns correspond to training and evaluation sets, respectively.}
    \label{tab:results}
    \vspace{-1em}
\end{table}


\section{Conclusion}



In this work, we introduce a novel loss based on Optimal Transport for enforcing translation equivariance. We show promising results on the task of self-supervised pitch estimation, with performances close to state-of-the-art.

Furthermore, equivariance to translations is at the core of various MIR tasks such as self-supervised  tempo~\cite{Gagnere2024}, key~\cite{kongSTONESelfsupervisedTonality2024, kongSKEYSelfsupervisedLearning2025}, chord or multi-pitch~\cite{Cwitkowitz2024} estimation. We therefore believe that our contribution may have useful applications beyond the scope of this paper.
In particular, variants such as Circular OT~\cite{rabinTransportationDistancesCircle2011} may be of particular interest for tasks such as key and chord estimation.

\section{Acknowledgments}

This work was funded by the European Union (ERC, HI-Audio, 101052978). Views and opinions expressed are however those of the author(s) only and do not necessarily reflect those of the European Union or the European Research Council. Neither the European Union nor the granting authority can be held responsible for them.

{
\bibliography{macros, bern, alain}}

%
%
%
%
%

\end{document}